\documentclass{article}

\pdfoutput=1

\usepackage{xcolor}
\definecolor{logobgcolor}{HTML}{FF5C33}

\usepackage{listings}
\newcommand{\setlistinglanguage}[1]{
	\lstset{
		backgroundcolor=,
		basicstyle=\small\ttfamily,
		belowcaptionskip=-1.25\baselineskip,
		breakatwhitespace=true,
		breaklines=false,
		captionpos=t,
		commentstyle=\color[gray]{0.5},
		escapeinside={\%*}{*)},
		extendedchars=true,
		frame=single,
		keepspaces=true,
		keywordstyle=\color{logobgcolor}\ttfamily,
		language={#1},
		morekeywords={*,...},
		numbers=left,
		numbersep=7pt,
		numberstyle=\tiny\color{black},
		rulecolor=\color{black},
		showspaces=false,
		showstringspaces=false,
		showtabs=false,
		stepnumber=1,
		stringstyle=\color{red},
		tabsize=2,
		title=\lstname,
	}
}
\setlistinglanguage{Java}

\usepackage{hyperref}

\title{A Modularity Bug in Java~8}
\author{Simon KRAMER\\[\jot]
		\texttt{simon.kramer@a3.epfl.ch}}
\date{December 31, 2016}

\begin{document}
\maketitle

\begin{abstract}
	We demonstrate a modularity bug in the interface system of Java~8 
		on the practical example of a textbook design of a modular interface for 
			vector spaces.
	Our example originates in our 
		teaching of modular object-oriented design in Java~8 to undergraduate students, 
			simply following standard programming practices and mathematical definitions.
	The bug 
		shows up as a compilation error and 
		should be fixed with a language extension 
			due to the importance of best practices (design fidelity).
			
	\medskip
	
	\noindent
	\textbf{Keywords:}
		component-based software construction, 
		interface specifications, 
		object-oriented programming,
		programming-language pragmatics
\end{abstract}

\section{Introduction}
	We demonstrate a modularity bug in the interface system of Java~8 \cite{Java8} 
		on the practical example of a textbook design of a modular interface for 
			vector spaces, which
				are very important for many computer applications of linear algebra.
	Our design is textbook in the sense that 
		we simply follow standard programming practices and mathematical definitions,
			as any mathematically-inclined teacher of object-oriented programming in Java would do.
	That is, 
		we construct our design from its standard algebraic components,
			thus ending up with not only \emph{a} but actually \emph{the} natural modular design (by the algebraic definition of vector spaces), which
				however---and to our great surprise and embarrassment to our students---fails compilation due to 
					an important expressiveness limitation of the Java~8 language.
	This limitation 
		resides in its (not so) generic type system and 
		should be remedied with an appropriate language extension,
			for the sake of 
				keeping best practices best and 
				saving our faces to our students of Java.

	\paragraph{Nota bene} 
		We do \emph{not} claim that 
			it be impossible to model vector spaces in Java~8;
				it \emph{is} possible.
		What we do claim is that 
			it is impossible to 
					specify with interfaces and 
					implement with corresponding classes 
				such spaces in Java~8 \emph{in the natural modular way,} whereby 
					we mean \emph{reflecting their algebraic structure with its definitional components} 
						(design fidelity).
	So the great modularity promise of component-based software construction with 
		object-oriented programming in Java 
			has not been fulfilled yet, even after more than 20 years. 
	In particular (counterexample), 
		it cannot be fulfilled in Java~8 when \emph{combining} such basic components as 
			a field of scalars (e.g., rational numbers) and 
			an additive group of vectors of such scalars so as  
				to form a vector space (e.g., over rationals).
	Unfortunately,
		not even the planned Java~9 module system \cite{Java9Modularity} 
			\begin{itemize}
				\item representing 
						``a fundamental shift to modularity as a first-class citizen for the whole Java platform'' and
				\item in which ``interfaces play a central role''
			\end{itemize}
			offers the required basic modularity,  
				perhaps because of prioritising  
					packages (coarser-grained modularity through import and export control) over  
					interfaces (finer-grained modularity through multiple-inheritance-based composition).

\section{The bug}
We demonstrate the modularity bug in the interface system of Java~8 by
	constructing our interface (specification) for vector spaces 
		bottom-up from its standard algebraic (interface) components, 
			combining them through multiple inheritance when needed.
Luckily, Java~8 offers multiple inheritance from interfaces.
Recall that from classes, Java~8 only offers single inheritance.
Our following interface components specify the algebraic operations up to our sought vector spaces, 
	however without stipulating their corresponding algebraic (equational) laws (constraints).
It would be nice if also this could become possible in a future release of Java in which 
	interfaces could carry such algebraic structure, 
		\emph{\`{a} la} algebraic specifications \cite{AlgebraicSpecifications}.
In particular,
	the commutativity of commutative rings is left implicit (unspecified) in the following corresponding interface (Page~\pageref{page:CR}).

Our interfaces are presented as divided up into non-problematic and problematic components.
The non-problematic interface components are meant to illustrate 
	important modular (algebraic) interface specifications that \emph{are} possible in Java~8
		(though unfortunately still without algebraic constraints).
The problematic interface components are meant to demonstrate 
	an important modular (algebraic) interface specification that is \emph{not} possible in Java~8, namely
		the one for vector spaces (our bug).
The frontier between the non-problematic and the problematic 
	is our empirical evidence for our claimed expressiveness limitation in Java~8 and 
	motivates our suggested language extension therefor.

\subsection{Non-problematic interfaces}
Our first (atomic) interface component for vector spaces is the following one for (commutative) additive semigroups.
These carry one binary operation, here called \emph{plus}, 
		specified with an explicit parameter \emph{right},
			with the corresponding parameter \emph{left} left implicit, as
				is typical in object-oriented programming languages.
(We can refer to this implicit parameter with 
	the Java-keyword \textcolor{logobgcolor}{\texttt{this}}.)
\newpage
	\begin{center}
		\texttt{algebra.AdditiveSemigroup}\\
		\begin{lstlisting}
package algebra;

interface AdditiveSemigroup <T> {

	T plus (T right);

}\end{lstlisting}
	\end{center}
A typical example of this strucure are 
	the natural numbers (naturals) with addition.
They actually form also a (commutative) multiplicative semigroup (and more), which 
	is our second atomic interface component for vector spaces.
	\begin{center}
		\texttt{algebra.MultiplicativeSemigroup}\\
		\begin{lstlisting}
package algebra;

interface MultiplicativeSemigroup <T> {

	T times (T right);

}\end{lstlisting}
	\end{center}
Our next (compound) interface component for vector spaces is the following one for additive monoids, which
	we obtain 
		through single inheritance from our interface component for 
			additive semigroups and 
		through the addendum of a getter method for 
			the additionally required additive neutral element.
	\begin{center}
		\texttt{algebra.AdditiveMonoid}\\
		\begin{lstlisting}
package algebra;

interface AdditiveMonoid		<T> 
	extends AdditiveSemigroup <T> {

	T getZero(); // the additive neutral element

}\end{lstlisting}
	\end{center}
As a typical example, the naturals form (also) an additive monoid.
Moreover, they form a multiplicative monoid, which 
	is similar to its additive counterpart.
	\begin{center}
		\texttt{algebra.MultiplicativeMonoid}\\
		\begin{lstlisting}
package algebra;

interface MultiplicativeMonoid		<T> 
	extends MultiplicativeSemigroup <T> {

	T getOne(); // the multiplicative neutral element

}\end{lstlisting}
	\end{center}
Our next, further compound interface component for vector spaces is the following one for additive groups, which
	we obtain 
		through single inheritance from our interface component for 
			additive monoids and 
		through the addendum of a getter method for 
			the additionally required additive inverse element.	
	\begin{center}
		\texttt{algebra.AdditiveGroup}\\
		\begin{lstlisting}
package algebra;

interface AdditiveGroup	 <T> 
	extends AdditiveMonoid <T> {

	T getAddInv(); // the additive inverse element

}\end{lstlisting}
	\end{center}
As a typical example, the integer numbers (integers) form an additive group.
Moreover, they form a commutative ring, which 
	we obtain through multiple (double) inheritance from 
		our (two) interface components 
			for additive groups on the one hand and 
			for multiplicative monoids on the other hand.
\label{page:CR}
	\begin{center}
		\texttt{algebra.CommutativeRing}\\
		\begin{lstlisting}
package algebra;

interface CommutativeRing			 <T> 
	extends AdditiveGroup				 <T>,
					MultiplicativeMonoid <T> {

}\end{lstlisting}
	\end{center}
Our next compound interface component for vector spaces is the following one for multiplicative groups, which
	we obtain 
		through single inheritance from our interface component for 
			multiplicative monoids and 
		through the addendum of a getter method for 
			the additionally required multiplicative inverse element.
	\begin{center}
		\texttt{algebra.MultiplicativeGroup}\\
		\begin{lstlisting}
package algebra;

interface MultiplicativeGroup	 <T> 
	extends MultiplicativeMonoid <T> {

	T getMultInv(); // the multiplicative inverse element

}\end{lstlisting}
	\end{center}
Our final non-problematic, further compound interface component for vector spaces is the following one for fields, which
	we obtain through
		\begin{itemize}
			\item multiple (double) inheritance from our (two) interface components
					for additive groups on the one hand and 
					for multiplicative groups on the other hand, and 
			\item the overriding 
					of the getter method for the multiplicative inverse element from the inherited multiplicative group  
					with an exception for division by zero.
					(Of course, we could define 
						a more general, algebraic exception instead of 
						getting by with the predefined arithmetic exception.)
Luckily, such an exceptional overriding is possible with Java~8 multiple inheritance.
		\end{itemize}
	\begin{center}
		\texttt{algebra.Field}\\
		\begin{lstlisting}
package algebra;

interface Field								<T> 
	extends AdditiveGroup				<T>,
					MultiplicativeGroup <T> {
					
	T getMultInv() throws ArithmeticException; // div by Zero!

}\end{lstlisting}
	\end{center}	
A typical example of a field are 
	the rational numbers (rationals) 
		with addition and multiplication, of course including 
			subtraction and division through 
				the inherited additive and the inherited multiplicative inverse elements, respectively.

\subsection{Problematic interfaces}
The desideratum of an interface for vector spaces in Java is 
	the one displayed in the upper part of Table~\ref{table:REQ}.
It is simply the one that the mathematical (algebraic) definition requires.
We obtain it 
	through multiple (double) inheritance from our (two) interface components 
		for an additive group of vectors on the one hand and 
		for a field of scalars on the other hand, and 
	through the addendum of a method for the scalar multiplication of vectors.
Very unfortunately,
	the desired interface produces the compilation errors displayed in the lower part of Table~\ref{table:REQ}, which
		cannot be remedied in Java~8 due to the limited expressiveness of its (not so) generic type system.
	\begin{table}[t!]
		\caption{\texttt{algebra.VectorSpace} (required version, but with compilation errors)}
		\begin{lstlisting}
package algebra;

interface VectorSpace		<Vector<Scalar>>
	extends AdditiveGroup <Vector<Scalar>>,
					Field								 <Scalar> {
					
	Vector<Scalar> timesScalar (Scalar s);

}\end{lstlisting}\setlistinglanguage{csh}
		\begin{lstlisting}
$javac algebra/VectorSpace.java 
algebra/VectorSpace.java:3: error: > expected
interface VectorSpace     <Vector<Scalar>>
                                 ^
algebra/VectorSpace.java:3: error: <identifier> expected
interface VectorSpace     <Vector<Scalar>>
                                        ^
algebra/VectorSpace.java:3: error: ';' expected
interface VectorSpace     <Vector<Scalar>>
                                          ^
algebra/VectorSpace.java:4: error: <identifier> expected
    extends AdditiveGroup <Vector<Scalar>>,
                                          ^
algebra/VectorSpace.java:4: error: ';' expected
    extends AdditiveGroup <Vector<Scalar>>,
                                           ^
algebra/VectorSpace.java:5: error: illegal start of type
            Field                <Scalar> {
                                          ^
algebra/VectorSpace.java:7: error: '(' expected
    Vector<Scalar> timesScalar(Scalar s);
    ^
algebra/VectorSpace.java:7: error: <identifier> expected
    Vector<Scalar> timesScalar(Scalar s);
                              ^
8 errors\end{lstlisting}
	\label{table:REQ}
	\end{table}
Even the simpler, less desirable, because type-wise less precise interface for vector spaces displayed in 
	the upper part of Table~\ref{table:AH} produces an intrinsic compilation error, which
		is displayed in the lower part.
	\begin{table}[t!]
		\caption{\texttt{algebra.VectorSpaceAH} (\emph{ad hoc} version, also with compilation error)}
		\setlistinglanguage{Java}
		\begin{lstlisting}
package algebra;

interface VectorSpaceAH	<Vector, Scalar>
	extends AdditiveGroup <Vector>,
					Field					<Scalar> {
					
	Vector timesScalar (Scalar s);

}\end{lstlisting}\setlistinglanguage{csh}
		\begin{lstlisting}
$javac algebra/VectorSpaceAH.java 
algebra/VectorSpaceAH.java:3: error: AdditiveGroup cannot \
be inherited with different arguments: <Vector> and <Scalar>
interface VectorSpaceAH   <Vector, Scalar>
^
1 error\end{lstlisting}
	\label{table:AH}
	\end{table}
(An even simpler interface would be a simplistic specification for vector spaces.)

Hence, multiple inheritance from interfaces with different generic types is problematic in Java~8.
This state of affairs has preempted---and still preempts, even in the future Java~9 \cite{Java9Modularity}---the 
		modularity of many interface specifications and thus 
			has worsened best practices in Java for more than 20 years.

\section{Conclusion}
We have demonstrated a modularity bug in the interface system of Java~8, which is due to 
	an important inadequacy in the Java generic type system, by 
		producing reproducible evidence in the form of 
			non-correctable compilation errors.
The type system of Java~8 should be generalised, in order to 
	allow the modular specification with interfaces of 
		such basic and important objects as vector spaces.
Currently, modular interface specifications only up to fields are possible.
It would also be very useful if 
	Java-interfaces allowed the stipulation of algebraic and other constraints on 
		their specified methods, at least to some extent.

\bibliographystyle{plain}

\end{document}